\newcounter{myctr}

\documentclass[pra, showkeys]{revtex4} \usepackage{latexsym}
\usepackage{amsfonts} \usepackage{amsmath,revsymb} 

\newcommand{\kt}[1]{\ensuremath{|#1\rangle}}
\newcommand{\br}[1]{\ensuremath {\langle #1|}}

\newcommand{\vp}{{\bf p}}

\newcommand{\SU}{{\rm SU}}

\newcommand{\HS}{{\mathcal{H}}}

\newcommand{\vc}[1]{{\bf #1}}

\begin{document}

\title{Dynamical Entanglement in Non-Relativistic, Elastic Scattering}

\author{N.L.~Harshman\footnote{Electronic address: harshman@american.edu}}
\affiliation{Department of Computer Science, Audio Technology and Physics\\
4400 Massachusetts Ave., NW \\ American University\\ Washington, DC 20016-8058}

\begin{abstract}
This article considers dynamical entanglement in non-relativistic particle scattering.  Three questions are explored: what kinds of entanglement occur in this system, how do global symmetries constrain entanglement, and how do the boundary conditions of scattering affect dynamical entanglement?  First, a simple model of scattering spin systems is considered, then the full system is discussed.
\end{abstract}

\keywords{Dynamical entanglement; scattering theory; Clebsch-Gordan methods}

\maketitle

\section{Introduction}

In scattering, the initial state of each beam (or the beam and the target) is represented by a density operator constructed from eigenvectors of the free Hamiltonian.  The total in-state is the direct product of these density matrices and is separable.  The interaction provides the mechanism for entanglement and the final state is generally not separable.  This paper studies this kind of dynamical entanglement in scattering.  This subject has recently received increased interest~\cite{mack02,lamata06b}.

In particular, this work investigates how entanglement is dynamically generated in elastic, non-relativistic, two-particle scattering processes.  The following questions will receive attention: (1) What kinds of entanglement occur in non-relativistic particle scattering?  Non-relativistic particles have continuous (e.g.,~momentum) and discrete (spin component) degrees of freedom, and therefore the state space has a richer structure and allows for different manifestations of entanglement. (2) How do conservation laws limit the dynamics of entanglement?  Since the dynamical evolution of the system must respect global Galilean invariance, not all entangling operations can arise through scattering processes. (3) How do the boundary conditions of scattering affect the dynamics of entanglement?  The in-state must be separable.  What other requirements on the interaction and on the in-state are necessary or sufficient so that entanglement is generated?

These questions do not have complete answers.  In what follows, techniques from symmetry representations in quantum mechanics, like Clebsch-Gordan methods, will be combined with the S-matrix formulation of scattering dynamics to give partial answers and uncover new avenues for further inquiry.  Only pure states will be considered.

\section{Scattering Spin Systems}

To develop intuition and some terminology, we first consider a simpler model of scattering.  We will neglect the momentum degrees of freedom of non-relativistic particles and consider only the effect of scattering on the spin degrees of freedom.  In other words, we can consider two spin systems interacting for some finite time interval by a rotationally-invariant interaction (the relevant global symmetry).  As will be discussed below, low-energy scattering by a central interaction approximates the model of scattering spin systems.

For simplicity, we consider the scattering interaction of two spin-1/2 systems, particle A and particle B.  The Hilbert space for two spin-1/2 systems is the tensor product of each particle's Hilbert space $\HS^{AB} = \HS^A\otimes\HS^B$.
Since the free systems before and after the interaction have rotational symmetry, each $\HS^i$ is the representation space for the unitary irreducible representation (UIR) $D^{(1/2)}$ of $\SU(2)$.  For the in-state and out-state, we will use the spin component basis, i.e.~
\begin{equation}
S_3^A\kt{\pm}_A = \pm(\hbar/2)\kt{\pm}_A.
\end{equation}
We assume that the two systems are initially in pure states of spin and that they have never interacted before.  The coordinates can always be chosen so that the initial state of particle A is $\kt{\phi_A^{in}}=\kt{+}_A$.  Then, except for an overall phase, the most general state for the second particle is
\begin{equation}
\kt{\phi_B^{in}}=\cos(\theta/2)\kt{+}_B + e^{i\phi/2}\sin(\theta/2)\kt{-}_B,
\end{equation}
where $0 \leq \theta \leq \pi$ and $0 \leq \phi \leq 4\pi$.
The total in-state $\kt{\phi^{in}}$ is the direct product
\begin{equation}\label{phiin}
\kt{\phi^{in}} = \cos(\theta/2)\kt{++} + e^{i\phi/2}\sin(\theta/2)\kt{+-},
\end{equation}
where, for example, $\kt{++} = \kt{+}_A\otimes\kt{+}_B$.  This is a stationary state of the free Hamiltonian before the scattering.

Scattering brings the spin degrees of freedom into contact with each other.  After the interaction is over, the new out-state $\kt{\phi^{out}}$ will again be a stationary state of the free Hamiltonian.  The dynamics of scattering can be represented by the unitary S-operator which transforms the in-state into the out-state $\mathsf{S}\kt{\phi^{in}}=\kt{\phi^{out}}$.  As a consequence of rotational symmetry, the S-operator must satisfy the relation
\begin{equation}\label{spincons}
[\mathsf{S},\vc{S}]= 0
\end{equation}
where $\vc{S}$ is the total spin operator with component operators $S_i = S_i^A\otimes\mathbb{I} + \mathbb{I}\otimes S_i^B$.
Since $\vc{S}$ is a vector operator and the S-matrix is a scalar operator, the Wigner-Eckart theorem tells us that the S-matrix can only be a function of total spin $s$.  Therefore, dynamics will be more transparent if we decompose the total Hilbert space into a direct sum of UIRs of the global symmetry $\HS^A\otimes\HS^A=\HS^{s=0}\oplus\HS^{s=1}$.  The in-state (\ref{phiin}) is transformed  into the total spin basis $\kt{s s_3}$ using the standard Clebsch-Gordan coefficients for the rotation group to get
\begin{equation}
\kt{\phi^{in}} = \cos(\theta/2)\kt{11} + \frac{1}{\sqrt{2}}e^{i\phi/2}\sin(\theta/2)\left(\kt{10} + \kt{00} \right).
\end{equation}

Unitarity constrains the S-operator and the S-matrix in the direct sum basis has the form
\begin{equation}\label{spinS}
\br{s s_3}\mathsf{S}\kt{s' s_3'} = e^{2i\delta_s}\delta_{ss'}\delta_{s_3s_3'},
\end{equation}
where $\delta_s$ are the phase shifts.  All possible dynamics in spin-1/2 scattering systems can be specified by two parameters $\delta_0 \in[0,\pi)$ and $\delta_1 \in[0,\pi)$.  Further, we expect that a global phase is irrelevant and so only the difference $(\delta_0 - \delta_1)\in(-\pi,\pi)$ should affect physical observations.

Using (\ref{phiin}) and (\ref{spinS}), the most general out-state (within the chosen coordinate system) is 
\begin{eqnarray}
\kt{\phi^{out}} &=& S\kt{\phi^{in}}\nonumber\\
& =& \cos(\theta/2)e^{2i\delta_1}\kt{11} + \frac{1}{\sqrt{2}}e^{i\phi/2}\sin(\theta/2)\left(e^{2i\delta_1}\kt{10} + e^{2i\delta_0}\kt{00}\right)\nonumber\\
& =& \cos(\theta/2)e^{2i\delta_1}\kt{++}\label{phiout}\\
&&+ \frac{1}{2}e^{i\phi/2}\sin(\theta/2)\left\{\left(e^{2i\delta_1}+e^{2i\delta_0}\right)\kt{+-} + \left(e^{2i\delta_1}-e^{2i\delta_0}\right)\kt{-+}\right\}.\nonumber
\end{eqnarray}
So for this simple model, four parameters $\{\phi,\theta,\delta_0,\delta_1\}$ are sufficient to describe every possible in-state (the first two parameters) and all dynamics arising from central forces (the second two).  We now turn to the three questions mentioned in the introduction.

What kinds of entanglement occur in non-relativistic particle scattering?  In this model, the global Hilbert space is finite dimensional and can be factored into the direct product of two Hilbert subspaces.  Therefore, the entropy of entanglement (EoE) provides an unambiguous measure of entanglement in pure states.  For the full case of momentum and spin degrees of freedom, the answer to this question is more complicated, as will be discussed below.

Calculating the EoE of (\ref{phiout}), we find
\begin{equation}\label{spinent}
E(\kt{\phi^{out}}) = 1 - \log_2\left[(1+x)^{(1+x)}(1-x)^{(1-x)}\right],
\end{equation}
where
\begin{equation}
x = \sqrt{ 1 - \sin^4(\theta/2)\sin^2(2\Delta\delta)}
\end{equation}
and $\Delta\delta = \delta_0 - \delta_1$.  The EoE (\ref{spinent}) has a maximum $E=1$ when $x=0$ and a minimum $E=0$ (no entanglement) when $x=1$.

How do symmetries limit the dynamics of entanglement?  In this case, rotational symmetry means that the S-matrix must have the highly-restricted form (\ref{spinS}).  From (\ref{spinent}), we see that the phase shift difference $\Delta\delta$ determines whether the dynamics can be entangling.  
In particular, we see that only for $\Delta\delta = \pm\pi/4$ can the S-operator lead to maximal entanglement.  Such operators have been called perfect entanglers~\cite{zhang02} because they are operators which can take an unentangled vector to a maximally-entangled vector.  

How do the boundary conditions of scattering affect the dynamics of entanglement?  The EoE of the in-state is zero by construction as a consequence of the assumption of inital independence of the particles.  For which initial states are there perfectly-entangling dynamics?  In other words, for which values of $\phi$ and $\theta$ are there solutions to $E(\kt{\phi^{out}})=1$?  The phase $\phi$ is unimportant, and only when $\theta = \pi$ is the in-state $\kt{\phi^{in}}=\kt{+-}$ maximally-entangleable.  This in-state has equal components in both partial waves.
Even then, maximal entanglement can only occur for $\Delta\delta = \pm\pi/4$.
In this case, the out-state is
\begin{equation}
\kt{\phi^{out}} = \omega\frac{1}{\sqrt{2}}(\kt{+-} \mp i\kt{-+})
\end{equation}
where there is an overall phase $\omega=\exp(i\phi/2)\exp(2i\delta_1)\exp(\pm i\pi/4)$.

The lessons of the toy model can be summarized as follows:  Enforcing global symmetry limits the set of perfectly-entangling dynamical operators and the sets of separable in-states that can be perfectly entangled.  For some in-states, no entanglement is possible for any dynamics.  Finally, only a limited set of maximally-entangled out-states are accessible via dynamical entanglement.

\section{The Full Problem}

Next we discuss the problem of entanglement in non-relativistic scattering.  We will attempt to show that all of the above observations also appear to be true for the full problem, although there are many complications and many other kinds of entanglement phenomena can occur.

The first step is to cast the mathematics in terms of the language of symmetry.  The symmetry group is now the Galilean group $\mathcal{G}$ (or, more properly, its quantum-mechanical covering group $\tilde{\mathcal{G}}$).  UIRs of $\tilde{\mathcal{G}}$ are associated with a single, free, non-relativistic particles~\cite{levy63}.  They are denoted $D^{(m,w,s)}$ and are characterized by invariant mass $m$, internal energy $w$, and spin $s$.  The representation space is denoted $\HS^{(m,w,s)}$.  Eigenkets of momentum and spin component $\kt{\vp,s_3(mws)}\in(\Phi^{(m,w,s)})^\times$ form a basis for a dense subspace $\Phi^{(m,w,s)}\subset\HS^{(m,w,s)}$ of well-behaved vectors (see Ref.~(\cite{rhs}) for details).
Unlike in the case of relativistic particles, the space $\Phi^{(m,w,s)}$ can be factorized into the direct product of momentum and spin degrees of freedom:
\begin{equation}
\Phi^{(m,w,s)}=\Phi^{(m,w)}\otimes\mathcal{H}^{(2s+1)}.
\end{equation}

It is assumed that a pure in-state is separable, i.e.~$\kt{\phi^{in}} = \kt{\phi^{in}_A}\otimes\kt{\phi^{in}_B}$
where $\kt{\phi^{in}_i}\in\Phi^{(m_i,w_i,s_i)}$.
The S-operator transforms $\kt{\phi^{in}}$ to $\kt{\phi^{out}}$, and if the scattering is elastic, then we have
\begin{equation}
\kt{\phi^{out}}\in\Phi^{(m_1,w_1,s_1)}\otimes\Phi^{(m_2,w_2,s_2)},
\end{equation}
but in general it will no longer be separable.
Because of global Galilean invariance, the S-matrix will have a very restricted set of non-zero matrix elements, just as in the spin system model.  To exploit this, we can use Clebsch-Gordan methods to separate the direct product of single-particle UIRs into a direct sum of UIRs labeled by the global invariants:
\begin{equation}
\Phi^{(m_1,w_1,s_1)}\otimes\Phi^{(m_2,w_2,s_2)}=\bigoplus_{M,W,j,\eta}\Phi^{(M,W,j)[\eta_1,\eta_2]},
\end{equation}
where the set $\{M,W,j\}$ are the labels of the UIR and are global Galilean invariants with ranges that depend on the sinle particle invariants~\cite{levy63}.  
The same set of invariants may appear more than once in the direct sum; $\eta_1,\eta_2$ label the degeneracy.  In standard analysis, these are chosen to be the total spin angular momentum $s$ and the orbital angular momentum $l$ and this process is called partial wave analysis.

The Clebsch-Gordan coefficients can be used to transform the wave functions between the direct product basis $\kt{\vp^A,s_3^A(m_Aw_As_A)}\otimes\kt{\vp^B,s_3^B(m_Bw_Bs_B)}$ and the UIR direct sum basis with spin-orbit coupling $\kt{\vp j_3 (MWj)[ls]}=\kt{\vp j_3 n}$.  In the direct sum basis, because of Galilean invariance, the S-matrix will have the form~\cite{goldberger64}.
\begin{equation}\label{galS}
\br{\vp j_3 n}\mathsf{S}\kt{\vp' j_3' n'} = N(M,W)\delta^3(\vp -\vp')\delta(W -W')\delta_{jj'}\delta_{j_3j_3'}\br{ls}|\mathsf{S}^{(Wj)}|\kt{l's'},
\end{equation}
where $N(M,W)$ is determined by the normalization of the basis kets and $\delta_{j_3j_3'}\br{ls}|\mathsf{S}^{(Wj)}|\kt{l's'}$ is the reduced matrix element that mixes spin and orbital angular momentum.

What kinds of entanglement occur in this kind of scattering?  A much wider variety, because when two non-relativistic particles interact, the total Hilbert space cab be decomposed into the direct product of four independent subspaces, two for each particle's spin and two for momentum.  Tracing out the degrees of freedom of only one of these constituant spaces could result in a mixed state, pointing toward entanglement.  Taking a trace over different combinations could lead to EoEs associated with two-particle spin entanglement, two-particle momentum entanglement, internal entanglement between one particle's spin and momentum, and mixed varieties.  Even in-states with independent particles could have entanglement present in the form of internal spin-momentum entanglement.

Just like the spin system model, where global rotational symmetry limits when and how dynamical entanglement occurs, global Galilean symmetry has consequences for the dynamics and entanglement: not all unitary transformations are possible, so not all interactions are entangling.
Further, not all separable in-states may be entangleable.  The interplay between the symmetry of the dynamics and the boundary conditions of scattering could lead to some states that do not become entangled by any interaction, others that entangle under a wide class of interactions, and exhanges between different kinds of entanglement.

We conclude with one simple example that connects the full problem back to the spin systems model.  In the case of central interactions, $l$ and $s$ are also separately conserved.  Then a more convenient basis to understand the dynamics of entanglement is the $l$-$s$ basis $\kt{\vp l_3 s_3 (MW)[ls]}=\kt{\vp l_3 s_3 \tilde{n}}$, and the S-matrix elements in this basis are~\cite{goldberger64}
\begin{equation}\label{galls}
\br{\vp l_3 s_3 \tilde{n}}\mathsf{S}\kt{\vp' l_3' s_3' \tilde{n}'} = N(M,W)\delta^3(\vp -\vp')\delta(W -W')\delta_{ll'}\delta_{l_3l_3'}\delta_{ss'}\delta_{s_3s_3'}e^{2i\delta(W)_{ls}}.
\end{equation}
If the in-state is sharply peaked around a particular total energy $W$, and only one orbital angular momentum partial wave contributes to the scattering appreciably (as in the low energy limit), then the spin-dependent part of the interaction can be separated from the momentum-dependent part and the toy model above becomes a  realistic approximation~\cite{harshman05b}.

However, this is only the simplest case.  Many new phenomena and types of entanglement wait to be explored.  There are complications, namely working with continuous degrees of freedom, but recent progress with Gaussian states makes these difficulties appear tractable.  

\section*{Acknowledgments}

The author would like to thank the Dipartimento di Fisica at the Universit\'a degli Studi di Trento, the U.S.-Italian Fulbright Commission, and American University for the support they provided while he was on a Fulbright Junior Lectureship in Trento.

\end{document}